\documentstyle[11pt,newpasp,twoside,epsf]{article}
\newcommand{\umin}{u_{\rm min}}
\newcommand{\te}{t_{\rm E}}

\newcommand{\re}{R_{\rm E}}

\newcommand{\cs}{$\chi^{2}\:$} 

\newcommand{\Am}{$A_{\rm max}\:$}
\newcommand{\tf}{$t_{\rm FWHM}$}

\reversemarginpar

\begin{document}
\title{Detecting Extra-Solar Planets via Microlensing at High Magnification}

\author{Nicholas Rattenbury}
\affil{Department of Physics, The University of Auckland, Auckland, New Zealand.}
\author{Ian Bond}
\affil{The Royal Observatory, Edinburgh, United Kingdom.}
\author{Jovan Skuljan}
\affil{Department of Physics and Astronomy, The University of Canterbury, Christchurch, New Zealand.}
\author{Phil Yock}
\affil{Department of Physics, The University of Auckland, Auckland, New Zealand.}

\begin{abstract}
Extra-solar planets can be efficiently detected in gravitational microlensing events of high magnification. High accuracy photometry is required over a short, well-defined time interval only, of order 10-30 hours. Most planets orbiting the lens star are evidenced by perturbations of the microlensing light curve in this time. Consequently, telescope resources need be concentrated during this period only. Here we discuss some aspects of planet detection in these events.
\end{abstract}
\section{Introduction}

The presence of a planet in the lens system of a high magnification microlensing event results in a small deviation from the single lens light curve. This deviation occurs near the peak of the microlensing event and is detectable in events of high magnification (Liebes 1964; Griest \& Safizadeh 1998). Heavier mass planets induce a larger perturbation, and planets closer to the Einstein ring of the main lens also produce larger perturbations. The Einstein ring radius, $\re$, and crossing time, $\te$, gives the typical spatial and temporal scales for microlensing events:

\[
\re \simeq 1.9 \sqrt{\frac{M_{\rm L}}{0.3M_{\sun}}}  \:\:\rm AU \:\:\: \te \simeq 16.6\sqrt{\frac{M_{\rm L}}{0.3M_{\sun}}} \:\: \rm days.
\label{eq:bigre}
\]

The observer-lens and lens-source distances are assumed to be 6kpc and 8 kpc respectively. The projected transverse velocity of the source is assumed to be 220 $\rm kms^{-1}$. The time of full-width at half maximum for a high magnification event is: $t_{\rm FWHM} = \frac{3.5\te}{A_{\rm max}} $, where \Am is the maximum amplification of the event. A high magnification event occurs when the source and lens stars are well aligned, i.e. when $A_{\rm max} \simeq \frac{\re}{\umin} \gg 1$.

\section{Critical Observation time and Detectability regions}
\begin{figure}
\plotone{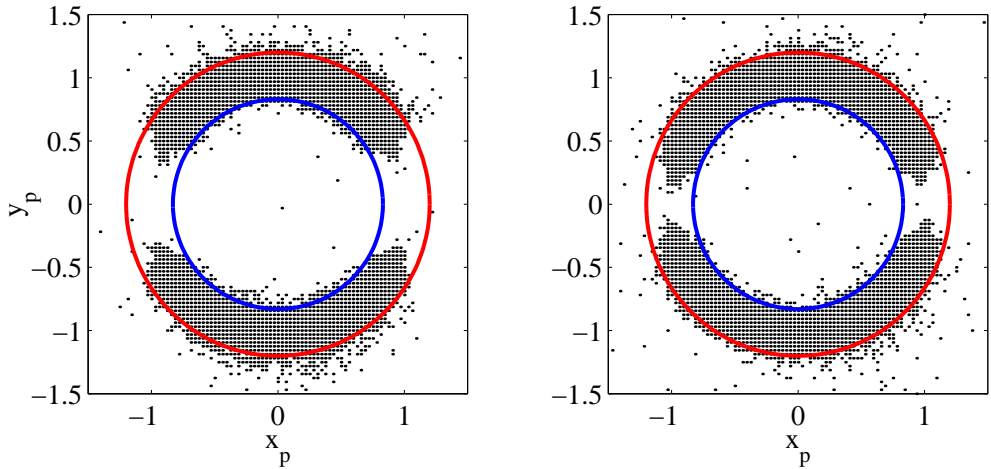}
\caption{\label{fig:detlims}Detection limit maps for an Earth-mass planet orbiting a $0.3M_{\oplus}$ lens star. The axes are in units of $\re$. The \emph{I}-band magnitude at maximum is 15, and the maximum amplification is 100. The left detection plot was generated using simulated observations over $[-\frac{1}{2}t_{\rm FWHM}, \frac{1}{2}t_{\rm FWHM}]$. The right plot used observations over the interval $[-t_{\rm FWHM}, t_{\rm FWHM}]$.}
\end{figure}

We define the time period around the peak amplification, \tf, as being the critical time to observe the event in order to detect any perturbations in the peak due to lens system planets. Simulations of high magnification microlensing events with a variety of planet-lens orbit radii and planet mass ratios were carried out on the University of Auckland cluster computer, \emph{Kalaka} (Rattenbury 2001). Each light curve thus generated was compared with a single lens light curve. If the difference between the perturbed light curve and the single lens light curve corresponded to a difference in \cs of 60, the perturbation was considered detectable by a 1-m class telescope. An example of the detection regions for an earth mass planet orbiting a $0.3M_{\odot}$ lens star is shown in Figure \ref{fig:detlims}. Simulated observations were only carried out in the time interval $+/- 0.5t_{\rm FWHM}$. Figure \ref{fig:detlims} shows the detection regions extending to either side of the Einstein ring. If observations are made just during the critical time \tf, then the data has the potential to contain perturbations in about $67\%$ of all possible lens-planet position angles.

\section{World-wide Microlensing Event Monitoring}
The major advantage of using high amplification microlensing events for the detection of extra-solar planets is the fact that the perturbation due to a lens system planet occurs near the peak of the event. The time of the event peak can be predicted well in advance. Telescope resources can therefore be concentrated during at least the critical time for detecting the planetary perturbations. In order to cover completely the critical time period for most events, a network of telescopes located around the world will be necessary.

\subsection{The MOA collaboration}
The Microlensing Observations in Astrophysics collaboration currently observes 16 Galactic bulge fields every night during the optimal southern hemisphere winter months using a 61cm reflector at the Mount John University Observatory. Microlensing events detected are reported via the Internet: \\ http://www.roe.ac.uk/$\sim$iab/alert/alert.html and by email alerts. Follow up observations of high magnification events are sought by other telescopes (Bond et al. 2002a). 

\section{An example: MACHO 98-BLG-35}

The MACHO collaboration alerted the microlensing community to a high magnification microlensing event which was subsequently followed up by several groups operating telescopes around the world. Analysis by the MPS and MOA groups found a perturbation in the peak of the light curve that may have been caused by a planet orbiting the lens star (Rhie et al. 2000). Re-analysis of the available data by the MOA collaboration using difference imaging yielded a perturbation that may have been caused by a low mass $(0.4 - 1.5M_{\oplus})$ planet (Bond et al. 2002). Continuous and accurate coverage of the critical time for this event allowed the reported analyses. The evidence of a low mass planet in a high magnification microlensing event prompted further investigation of the abilities and limitations of the technique.

\section{Capabilities of the High Magnification Technique}
\begin{figure}
\plottwo{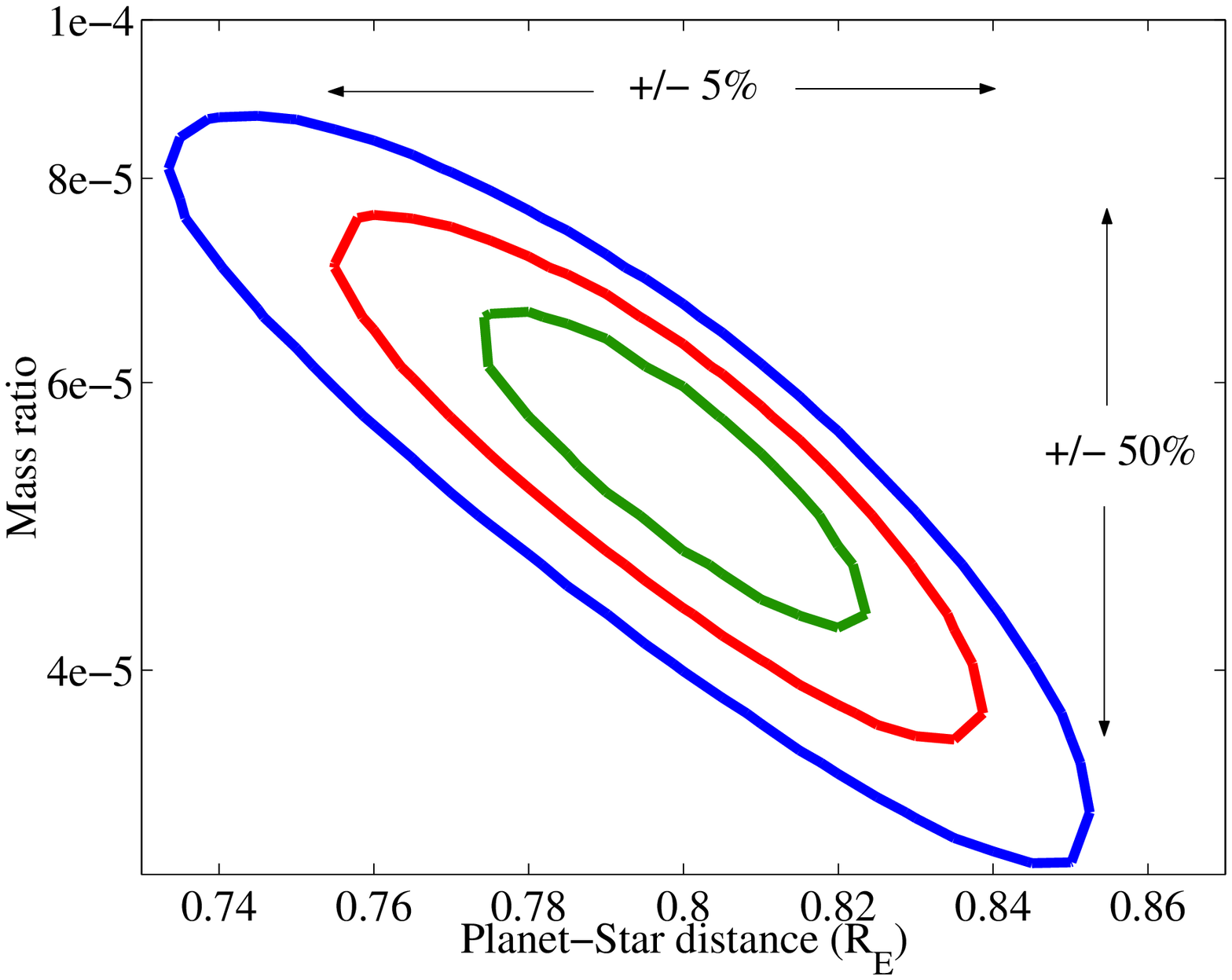}{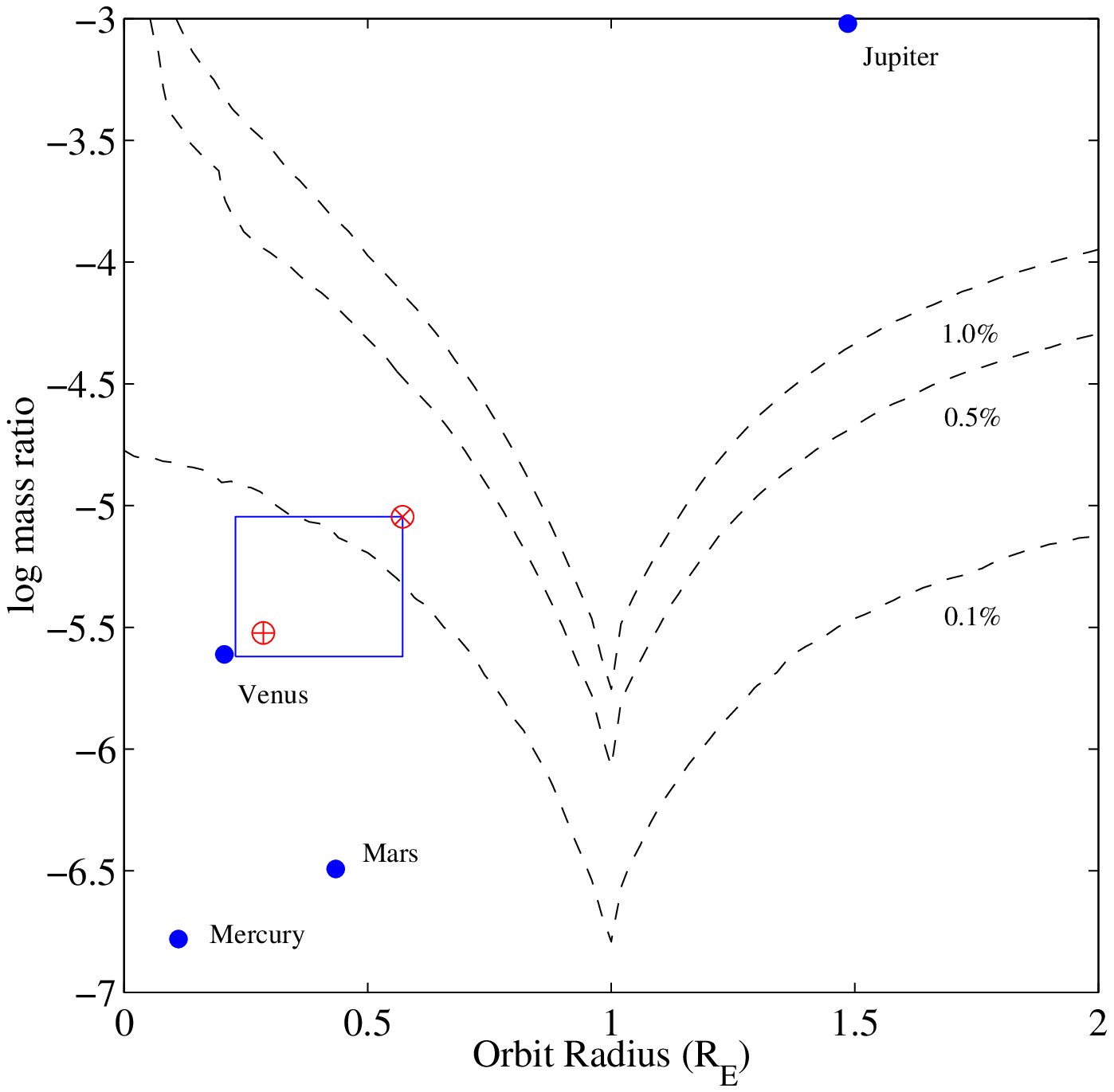}
\caption{\label{fig:double}(a) 1-,2- and $3\sigma$ contours for a planet at orbit radius and mass of $(a,\epsilon) = (0.8\re,5\times 10^{-5})$. (b) Mass-orbit radius detection limits for the high magnification technique for three values of peak light curve accuracy. The mass ratio and Einstein ring radius are for a solar mass lens star. The simulated light curves were computed over the time interval $[-2t_{\rm FWHM}, 2t_{\rm FWHM}]$, with $A_{\rm max} = 100$.}

\end{figure}

The detection regions for Earth, Neptune and Jupiter mass planets for a variety of microlensing situations were computed and are given in Rattenbury et al. (2002). The planet orbit radius and mass can be determined to within $\pm 5\%$ and $\pm 50\%$, respectively in typical cases, see Figure \ref{fig:double}a. Further simulations show that multiple planet systems can be easily modelled, provided that the position angle between planets exceeds about $20^{\circ}$. The finite source star size is inherent in the inverse ray shooting technique. Source star spots are not likely to be reasons for a false planetary perturbation, provided a caustic crossing does not occur. While low mass planets are detectable using this technique, planets in the habitable zone are probably not detectable, see Figure \ref{fig:double}b.

\section{Conclusions}
The observation of high magnification microlensing events is an effective method for the detection of extra-solar planets. The necessity of only observing during the critical time around the event peak enables efficient use of telescope scheduling and resources. Terrestrial mass planets are detectable using this technique, in regions bounding the Einstein ring of the lens system. For the most likely lens star mass, this distance is about $\simeq 2$AU. Probing the distribution of light mass planets at these distances around the host star will complement the current understanding of planet formation and distribution. This technique requires continuous, high accuracy photometry of the event around the peak. A network of 1 - 2 m class telescopes situated around the world will be able to perform the required observations. Future events are expected to yield initial statistics on the abundance of terrestrial mass planets.

NJR thanks the Graduate Research Fund of the University of Auckland, the LOC and the MOA project for financial assistance.

\end{document}